\documentclass[usenatbib]{mn2e}
\usepackage{graphicx}
\usepackage{amsmath}

\def\mnras{\rm MNRAS}

\def\gax{\mathrel{\raise.3ex\hbox{$>$}\mkern-14mu\lower0.6ex\hbox{$\sim$}}}
\def\lax{\mathrel{\raise.3ex\hbox{$<$}\mkern-14mu\lower0.6ex\hbox{$\sim$}}}
\def\gtorder{\mathrel{\raise.3ex\hbox{$>$}\mkern-14mu
             \lower0.6ex\hbox{$\sim$}}}
\def\ltorder{\mathrel{\raise.3ex\hbox{$<$}\mkern-14mu
             \lower0.6ex\hbox{$\sim$}}}

\begin{document}

\title [Transients Obscured by Dusty Disks]
   {Transients Obscured by Dusty Disks}

\author[C.~S. Kochanek]{ 
    C.~S. Kochanek$^{1,2}$
    \\
  $^{1}$ Department of Astronomy, The Ohio State University, 140 West 18th Avenue, Columbus OH 43210 \\
  $^{2}$ Center for Cosmology and AstroParticle Physics, The Ohio State University,
    191 W. Woodruff Avenue, Columbus OH 43210 \\
   }

\maketitle

\begin{abstract}
Dust absorption is invoked in a number of contexts for hiding a 
star that has survived some sort of transient event from view.
Dust formed in a transient is expanding away from the star and,
in spherical models, the mass and energy budgets implied by a
high optical depth at late times make such models untenable.  
Concentrating the dust in a disk or torus can in principle hide
a source from an equatorial observer using less mass and so
delay this problem. However, using axisymmetric dust radiation transfer
models with a range of equatorial dust concentrations, we find
that this is quite difficult to achieve in practice.  The polar
optical depth must be either low or high to avoid scattering
optical photons to equatorial observers. Most of the emission
remains at wavelengths easily observed by JWST, and the equatorial
brightness is reduced by at most a factor of $\sim 2$ compared to
isotropic emission even for equatorial (visual) optical depths
of $10^3$.  It is particularly difficult to hide a
source with silicate dusts because the absorption feature 
near $10\mu$m frequently leads to the emission being concentrated
just bluewards of the feature, near $8\mu$m.  
\end{abstract}


\section{Introduction}

There are several classes of transients in which dust forms and obscures
the surviving progenitor or is argued to form in order to explain the
apparent absence of a surviving progenitor at optical or near-IR wavelengths.  In
the first category are eruptions such as $\eta$ Car (e.g., \citealt{Humphreys1994}) 
and stellar mergers such as V838~Mon (e.g., \citealt{Bond2003}).  
In the second category are debated
transients such as the supernova impostors like SN~1997bs (e.g., \citealt{vandyk2000},
\citealt{Adams2015}) and 
SN~2008S (e.g., \citealt{Prieto2008}, \citealt{Thompson2009}, \citealt{Kochanek2011},
\citealt{Adams2016}) 
or the proposed failed supernovae NGC~6946-BH1 (\citealt{Gerke2015}, \citealt{Adams2017},
\citealt{Basinger2021}).  

With the constraining powers of ground based observatories, Hubble (HST)
and Spitzer (SST) Space Telescopes
 it is feasible to hide stars with the luminosities of these progenitor
systems from detection in nearby galaxies ($\ltorder 10$~Mpc)
with spherical shells of dusty ejecta.  The
optical depth and ejecta radius must simply be made large enough to
push the escaping emission to long enough wavelengths to evade
detection:  sufficiently absorbed in the optical and near-IR to be invisible 
or lost amid the overlapping sea of red giants, and with
dust emission cold enough to slip under the sensitivity limits of 
(generally) warm SST 
at $3.6$ and $4.5\mu$m.  Hiding the emission will become much more 
challenging in the era of JWST since it can easily detect such 
sources out to $\sim 25\mu$m with vastly better angular resolution
for avoiding confusion with either other stars or diffuse emission.

The combination of near-IR and bluer mid-IR observations generally
rule out sources obscured by dust forming in an on-going wind.  
Dust rapidly forms once temperatures drop below the condensation
temperature so, combined with the expansion, the optical depth of
a dusty wind is concentrated near its base.  Thus, a high optical
depth wind immediately converts the emission from the central source
into hot ($T_d \sim 1000$~K) dust emission.  It then takes extremely
high optical depths to absorb these photons further out in the wind
and shift the peak of the escaping emission beyond $5\mu$m. 

Colder dust emission is most easily achieved by forming dust in
material ejected in the transient. Dust forms and is initially
hot, but then becomes cooler as the ejecta moves outwards.
Dust growth ceases shortly after it forms, because the 
collisional growth rates are dropping as $r^{-2} \propto t^{-2}$ 
due to the expansion.  The dust opacity
is then constant, and the mean optical depth must drop as 
$t^{-2}$ assuming a constant expansion velocity and mass
conservation.  The effective optical depth could drop more 
rapidly than $t^{-2}$ if the expanding shell starts to 
fragment due to instabilities, leading to the radiation escaping
through lower optical depth channels in the ejecta.

The observed spectral energy distribution (SED) of the source can 
be used to constrain the luminosity ($L_*$) and 
temperature ($T_*$) of the central source,
the temperature of the dust ($T_d$) and
the visual optical depth $\tau$ of the shell.\footnote{Optical
depths in the text are all visual band ($0.55\mu$m) effective
optical depths $\tau = (\tau_a(\tau_s+\tau_a))^{1/2}$ given
the absorption and scattering optical depths $\tau_a$ and
$\tau_s$.  The effective optical depth takes into account the
extra path length created by the scattering. For the disk models,
the polar and equatorial visual effective optical depths are $\tau_p$
and $\tau_e$.}  The detailed
structure of the SED depends on the composition 
of the dust through the wavelength-dependent
opacities, and inferences about the mass of the ejecta depend
on the dust-to-gas mass ratio.  To order of magnitude, the
grain temperature is set by the available flux,
$\sigma T_d^4 = L_*/16 \pi r^2$, so constraints on the
overall luminosity and the dust temperature provides an
estimate of the dust radius $r$ which leads to an estimate
of the expansion velocity $v_e$ assuming $r= v_e \Delta t$
after elapsed time $\Delta t$.  Given an estimate of the
visual opacity $\kappa$, the SED model also provides an
estimate of the ejecta mass and kinetic energy.  For a thin
shell $\tau = M_e \kappa/4 \pi r^2$, so
\begin{equation}
    M_e = { 4 \pi r^2 \tau \over \kappa }
       = { L_* \tau \over 4 \kappa \sigma T_d^4}
\end{equation}
and
\begin{equation}
    E_e = { 2 \pi r^4 \tau \over \Delta t^2 \kappa }
       = { L_*^2 \tau \over 128 \Delta t^2 \kappa \sigma^2 T_d^8}.
\end{equation}
For the debated systems with non-detections, the question 
then becomes whether the velocities, masses and energies
required to make the dust cold enough to avoid detection
are plausible.  Time will also tell, since in any model
dependent on an expanding shell, the shell eventually becomes
transparent.

Where quantitative models are made for the SEDs of these sources,
they use spherically symmetric models, principally based on the
dust radiation transfer code {\tt DUSTY} (\citealt{Ivezic1997}, \citealt{Ivezic1999}). 
 Such models have two
shortcomings.  First, there is no reason the ejecta needs to be
spherically symmetric.  In particular, there are invocations of
models with more equatorial than polar material with the
observer near the equatorial plane as a means of evading the
limits implied by the spherical SED models (e.g., \citealt{Kashi2017},
\citealt{Andrews2021}, \citealt{Bear2022}).  Second, there is
no reason that the density distribution should remain homogeneous.
Expanding shells tend to develop instabilities, instabilities lead to
clumping of the material, and this produces lower optical depth channels
through which radiation can escape.  If instabilities develop,
the effective optical depth will drop faster than the 
$\tau \propto t^{-2}$ drop of the mean optical depth.   

In this paper we explore the first question, the effect of deviations 
from spherical symmetry, by examining the SEDs produced by axisymmetric
dusty shells with large differences between the equatorial and polar
optical depths.  For the optical and
near-IR light, the primary effect is that the polar dust can scatter
emission that would be absorbed if trying to directly escape 
along the equator into the line of sight of an equatorial observer.
In the mid-IR, a polar observer sees both less obscured direct emission
from the central source and dust emission with a broad range of
temperatures from the highly obscured regions.  Naively, the
equatorial observer sees reduced and colder emission.  We describe the
model used in \S2, present the results in \S3 and discuss their
consequences in \S4.

\section{Model}

\begin{figure}
\centering
\includegraphics[width=0.50\textwidth]{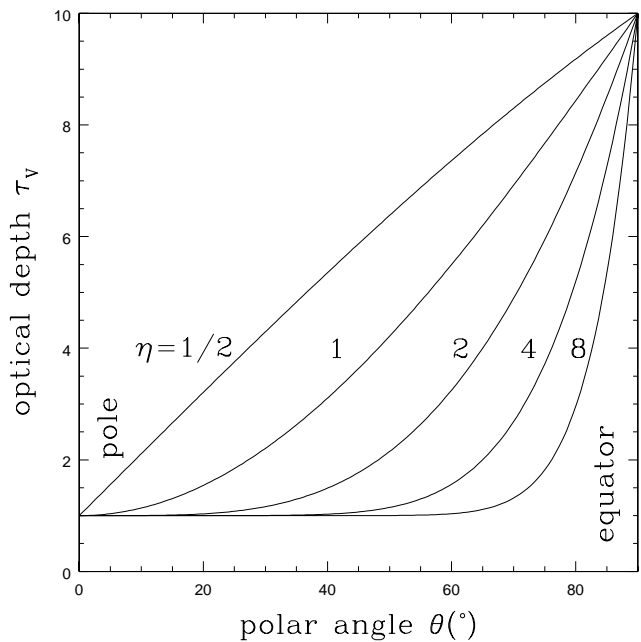}
\caption{
  Examples of the polar angle dependent optical depth produced by
  Eqn.~\protect\ref{eqn:density} with an equatorial optical depth of
  $\tau_e=10$, a polar optical depth of $\tau_p=1$ and exponents of
  $\eta=1/2$, $1$, $2$, $4$ and $8$.
  }
\label{fig:density}
\end{figure}

We used the {\tt RADMC-3D} (\citealt{Dullemond2012}) Monte Carlo dust radiation transfer package to
carry out the calculations in axisymmetry with reflection symmetry about the equator.  We
used the disky dust density model from \cite{Ueta2003},
$$
   \rho_d = 
   { \left[ \tau_p +  \left(\tau_e-\tau_p\right)  
   \left( 1 - \left|\cos \theta \right|\right)^\eta \right]
   \over R_{in} \kappa } 
    \left( 1  - { R_{in} \over R_{out}} \right)^{-1}
    \left( { R_{in} \over R} \right)^2
     \label{eqn:density}
$$
where $\tau_p$ and $\tau_e$ are the optical depths at the pole and the equator
($\tau_e > \tau_p$),
$R_{in}$ and $R_{out}$ are the inner and outer edges of the dust shell, and 
$\eta$ controls the concentration of the dust towards the equatorial plane,
as illustrated in Fig.~\ref{fig:density}. Larger
values of $\eta$ more strongly concentrate the dust towards the equator.  
We normalize the optical depths at V-band ($0.55\mu$m) based on the
effective absorption opacity
$\kappa = ( \kappa_a (\kappa_a + \kappa_s))^{1/2}$ 
given the absorption $\kappa_a$ and scattering $\kappa_s$ opacities.   
The effective absorption takes into account the increase in absorption
due to scattering lengthening the distance a photon travels.
The mass of the shell is
\begin{equation}
    M_{ej} = { 4 \pi R_{in} R_{out} \tau_e \over f_d\kappa }
          { 1 + \eta \tau_p/\tau_e \over 1 + \eta }
\end{equation}
where $f_d \simeq 0.005$ is the dust mass fraction in the shell.  
If $\tau_p/\tau_e \ll 1$, then the shell mass is $(1+\eta)^{-1}$ less than a 
spherical shell with the 
equatorial optical depth $\tau_e$ -- the dust has to be very equatorially 
concentrated (large $\eta$) before there is a large change in the required mass.
If the expansion velocity is $v_e$ and the elapsed time is $\Delta t$ so
that $R_{in}= v_e \Delta t$, the ejecta mass is
\begin{eqnarray}
    M_{e} &= &6.3 
      { \tau \over 100 } { R_{out} \over R_{in} }
       \left( { v_e \over 10^3~\hbox{km/s} }
        { \Delta t \over 10~\hbox{yr} }\right)^2
       \left( { 100~\hbox{cm}^2/\hbox{g} \over f_d \kappa }\right) \nonumber \\
       &&\times { 1 + \eta \tau_p/\tau_e \over 1 + \eta } M_\odot 
\end{eqnarray}
and the kinetic energy, assuming everything has a velocity of $v_e$, is
\begin{eqnarray}
    K_e &= &0.062
      { \tau \over 100 } { R_{out} \over R_{in} }
       \left( { v_e \over 10^3~\hbox{km/s} } \right)^4
        \left( { \Delta t \over 10~\hbox{yr} }\right)^2
       \left( { 100~\hbox{cm}^2/\hbox{g} \over f_d \kappa }\right) \nonumber \\
       &&\times{ 1 + \eta \tau_p/\tau_e \over 1 + \eta }~\hbox{FOE}
\end{eqnarray}
where $1~\hbox{FOE} = 10^{51}$~erg is the characteristic energy of
a supernova.  Assuming the velocity is the same for all material
is equivalent to assuming the shell thickness is due to the duration
of the event rather than a spread in velocity.  If we instead 
linearly increase the velocity with shell radius to explain 
the thickness, the kinetic energy would increase by 
$(R_{in}^2+R_{in}R_{out}+R_{out}^2)/3R_{in}^2$ or a factor
of $7/3$ for $R_{out}= 2R_{in}$.    

\begin{figure}
\centering
\includegraphics[width=0.50\textwidth]{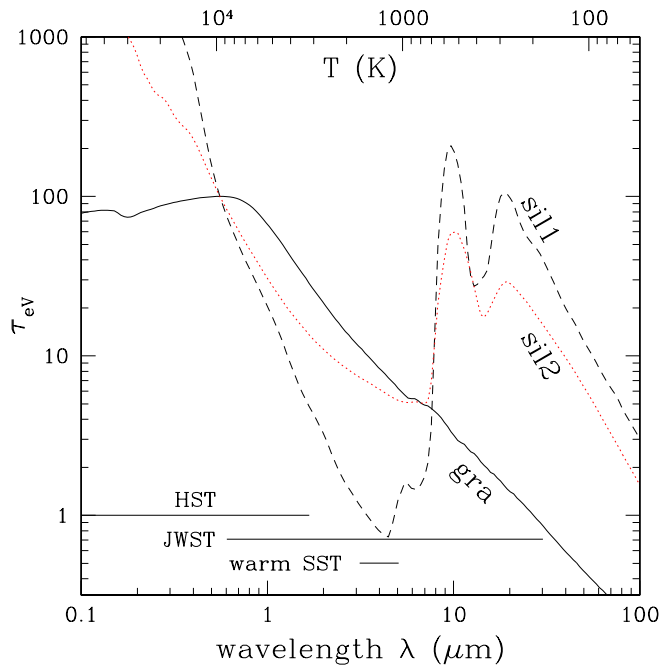}
\caption{
  The wavelength dependence of the effective optical depth for the 
  graphitic gra (black solid) and silicate sil1 (black dashed) and
  sil2 (red dotted) dust models 
  normalized to have an effective visual ($0.55\mu$m) 
  optical depth of $\tau=100$. The
  upper axis gives the temperature corresponding 
  to the mean photon energy ($T=h c/2.7 k \lambda$).  
  Horizontal bars at bottom show the wavelength ranges spanned
  by HST, warm SST and JWST.
  }
\label{fig:opdepth}
\end{figure}

We used a log normal grain size distribution with a mean radius of $0.1\mu$m and 
a width of $0.05\mu$m (i.e., 0.3~dex).  We generated the opacity tables with 
{\tt OpTool} (\citealt{Dominik2021}) and their default Distribution of Hollow 
Spheres (DHS, \citealt{Min2005}) opacity model.  We used the pyroxene 
silicate dust model (model ``sil1'', \citealt{Dorschner1995}) and the \cite{Draine2003} 
graphitic (model ``gra'') and astrosilicate (model ``sil2'') dust models.  
Fig.~\ref{fig:opdepth} shows the wavelength dependence of the effective 
absorption optical depth for the three models normalized to a visual
optical depth of $\tau=100$. One important difference between them
for the results is that graphitic dusts lack the strong silicate absorption 
feature at $10$-$20\mu$m.  One reason for including the two silicate models
is to explore the role of this feature, where the sil1 model has a peak
mid-IR opacity greater than its V band opacity, while the sil2 model has
a weaker peak.  The other important difference is that the silicate dusts
have much higher scattering opacities.  At $0.55\mu$m, the absorption,
scattering and effective opacities are
$(\kappa_a,\kappa_s,\kappa)=(380,8700,1900)$, $(2600,9300,5600)$, and 
$(93000,39000,110000)$~cm$^2$/g for the sil1, sil2 and gra models, respectively.      
Scattering is important only for the UV and optical -- it becomes
steadily less important relative to absorption moving into the infrared. 
Fig.~\ref{fig:opdepth} also shows the wavelength ranges covered by 
HST (WFC3/IR not NICMOS), warm SST
and JWST to illustrate the wavelength ranges where 
observations are feasible.
 
For these experiments, we model the central source based on the progenitor of
NGC~6946-BH1.  The central star is a black body treated as a point source 
with luminosity $L_* = 3 \times 10^5 L_\odot$ and temperature $T_*=4000$~K.
We include dust scattering, but {\tt RADMC-3D} is restricted to isotropic 
scattering for calculations in two dimensions unless doing a full scattering 
analysis with 
polarization.  We used the simpler, faster mode with isotropic scattering.
We used 100 radial zones between $R_{min}$ and $R_{max}=2 R_{min}$ and 90
angular zones between the equator and the pole.  The dust temperature was
set using $10^6$ photon packets, ten times the {\tt RADMC-3D} default and
the SEDs were generated using its default parameters.  

We used an expansion velocity of $v_e = 10^3$~km/s and considered expansion
times of $\Delta t = 1$, $3$, $5$, and $10$~years, leading to 
$R_{in}=632$, $1054$, $2108$, and $4216$~AU ($10^{15.98}$, $10^{16.20}$,
$10^{16.50}$ and $10^{16.80}$~cm).  Only the radius is actually relevant,
and these could just as well represent a system with $v_e=200$~km/s
expanding for $15$, $25$, $50$ and $100$~years.  Similarly, changes in
distance or time can also be viewed as changes in luminosity since the
emission is really controlled by $L_*/r^2$.  
We ran models with polar optical depths of $\tau_p=0.1$, $1.0$ and $3.0$ and 
equatorial optical depths of $\tau_e = 10$, $30$, $100$, $300$ and $1000$ over 
the range equatorial concentration power law indices from $\eta = 0.5$ to $8.0$ 
(see Fig.~\ref{fig:density}).  For comparison, we also ran 
spherically symmetric models at each optical depth.

\begin{figure*}
\centering
\includegraphics[width=1.00\textwidth]{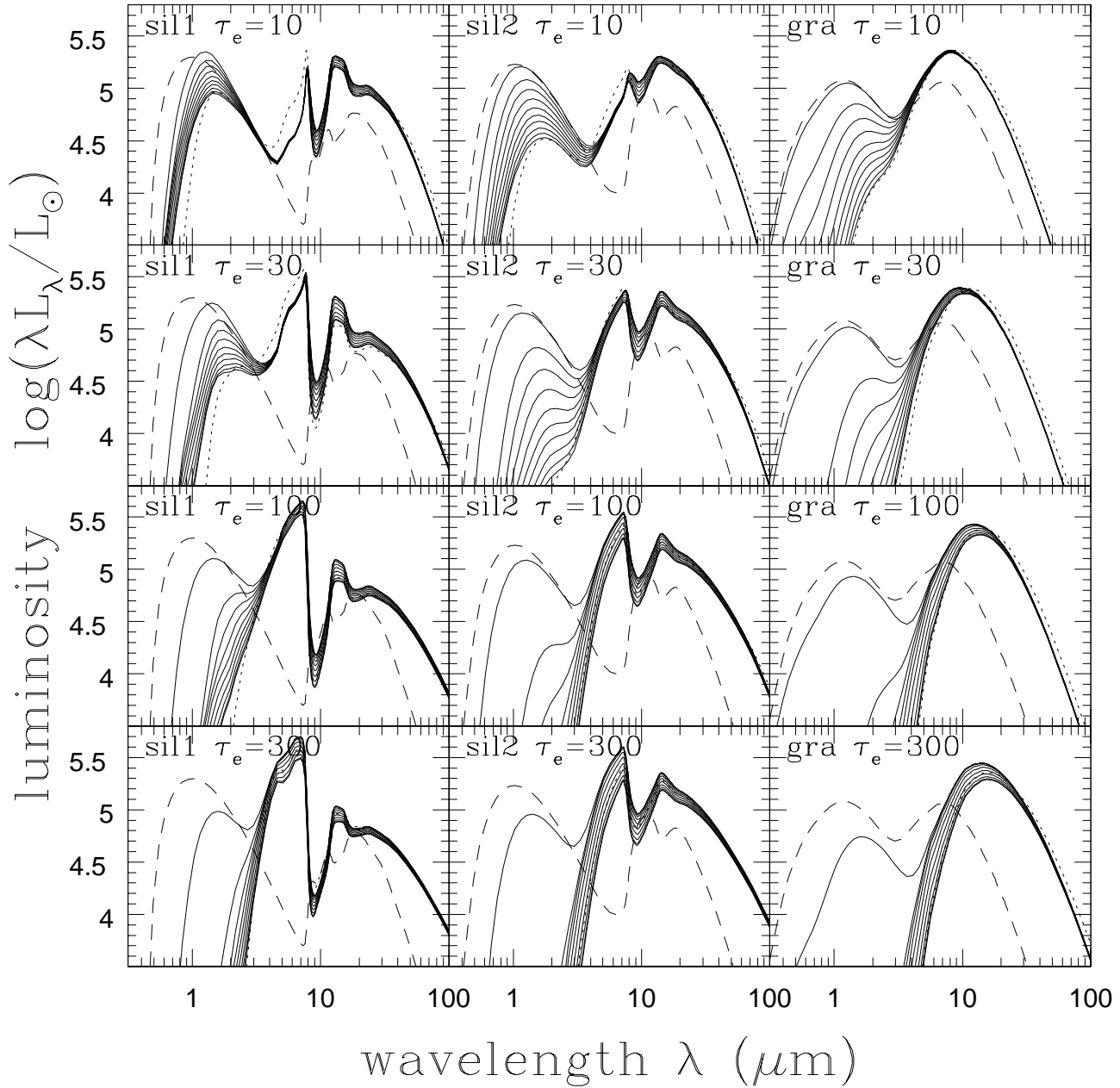}
\caption{
  Spectral energy distributions for models with $\eta=1/2$, $\Delta t = 3$~yr,
  polar optical
  depth $\tau_p=1$ and equatorial optical depths of $\tau_e=10$ (top),
  30 (middle), 100 (middle) and 300 (bottom) for the sil1 (left),
  sil2 (middle) and graphitic (right) dusts.  The solid lines are the SEDs of the 
  disky models viewed from
  $0^\circ$ (pole-on, most optical emission) to $90^\circ$ (edge on, least
  optical emission) in increments of $10^\circ$.  The dashed (dotted) line 
  is the SED for a spherical model with an optical depth of $\tau_p$ ($\tau_e$).  
  }
\label{fig:modfh}
\end{figure*}

\begin{figure*}
\centering
\includegraphics[width=1.00\textwidth]{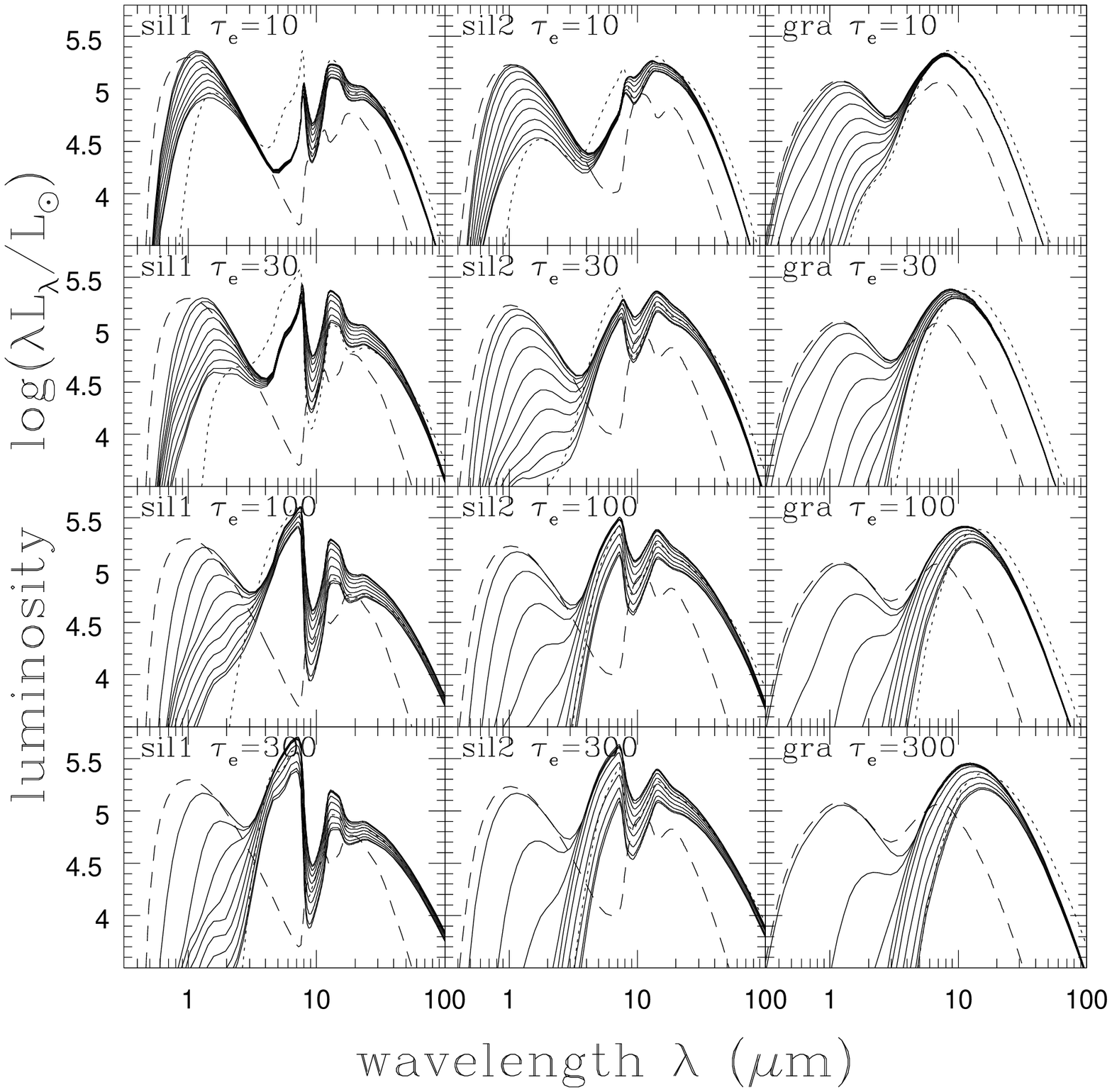}
\caption{
  Same as Fig.~\protect\ref{fig:modfh} but for $\eta=1$.
  }
\label{fig:modf1}
\end{figure*}

\begin{figure*}
\centering
\includegraphics[width=1.00\textwidth]{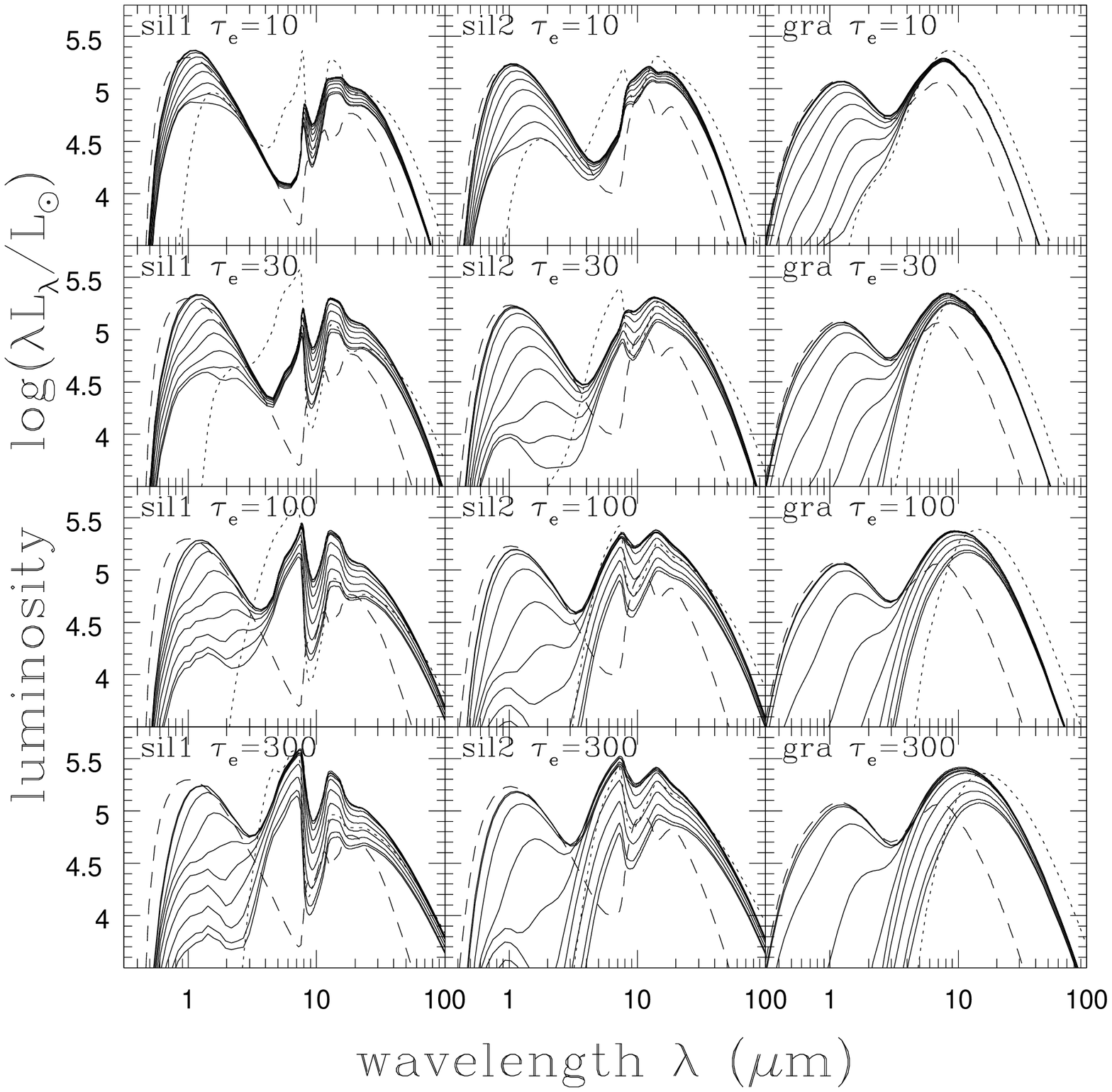}
\caption{
  Same as Fig.~\protect\ref{fig:modfh} but for $\eta=2$.
  }
\label{fig:modf2}
\end{figure*}

\begin{figure*}
\centering
\includegraphics[width=1.00\textwidth]{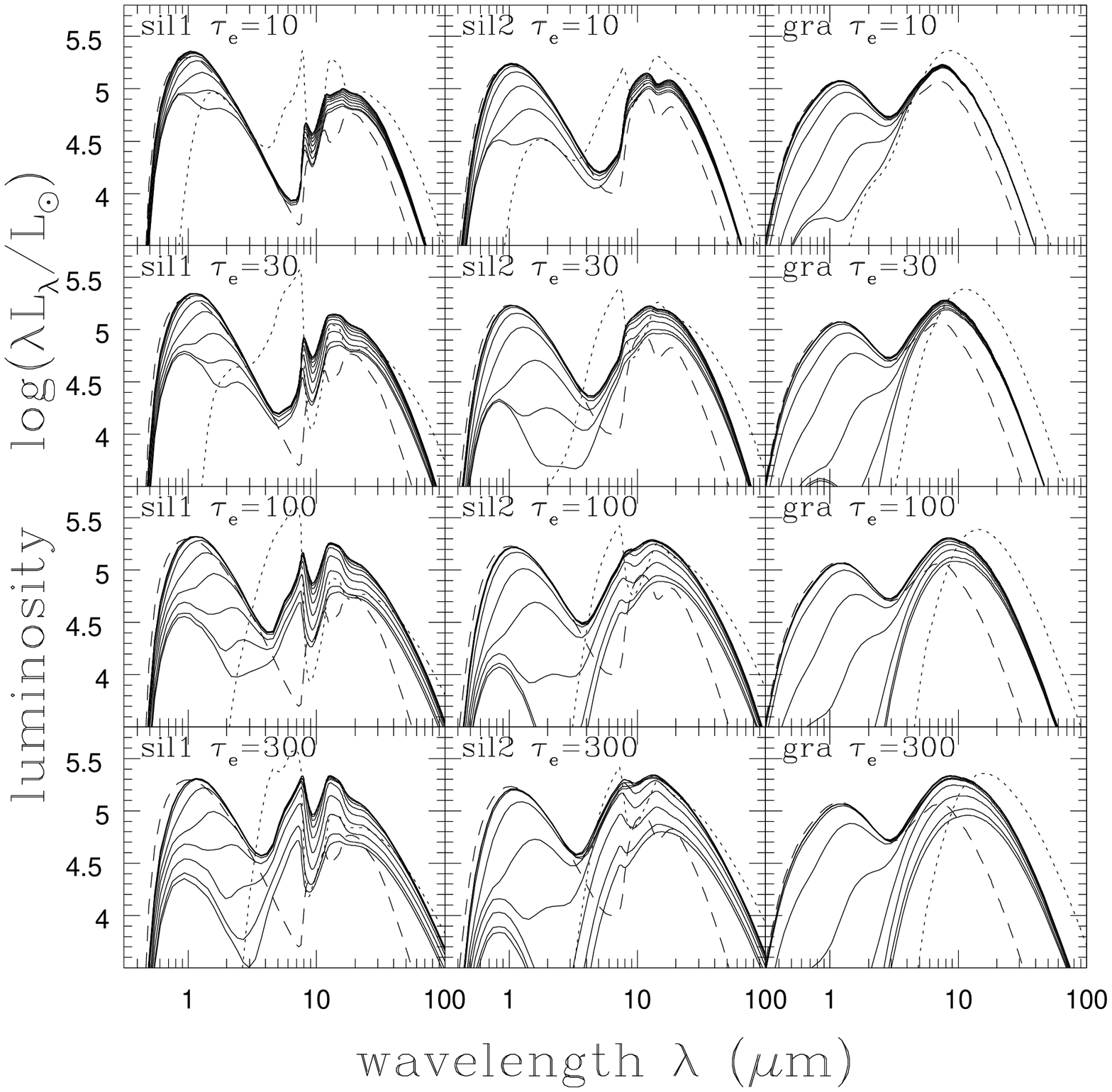}
\caption{
  Same as Fig.~\protect\ref{fig:modfh} but for $\eta=4$.
  }
\label{fig:modf4}
\end{figure*}

\begin{figure*}
\centering
\includegraphics[width=1.00\textwidth]{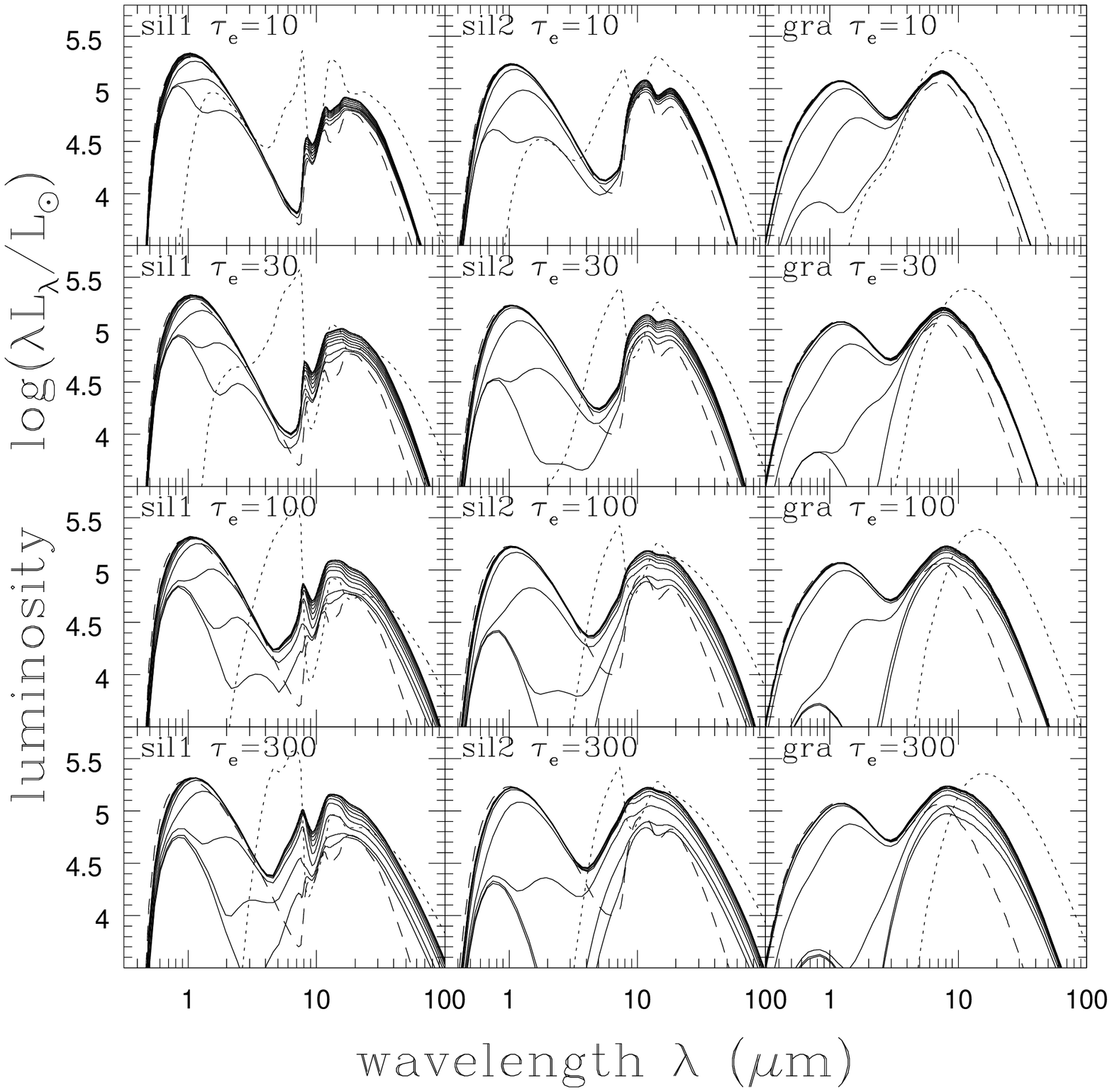}
\caption{
  Same as Fig.~\protect\ref{fig:modfh} but for $\eta=8$.
  }
\label{fig:modf8}
\end{figure*}

\begin{figure*}
\centering
\includegraphics[width=1.00\textwidth]{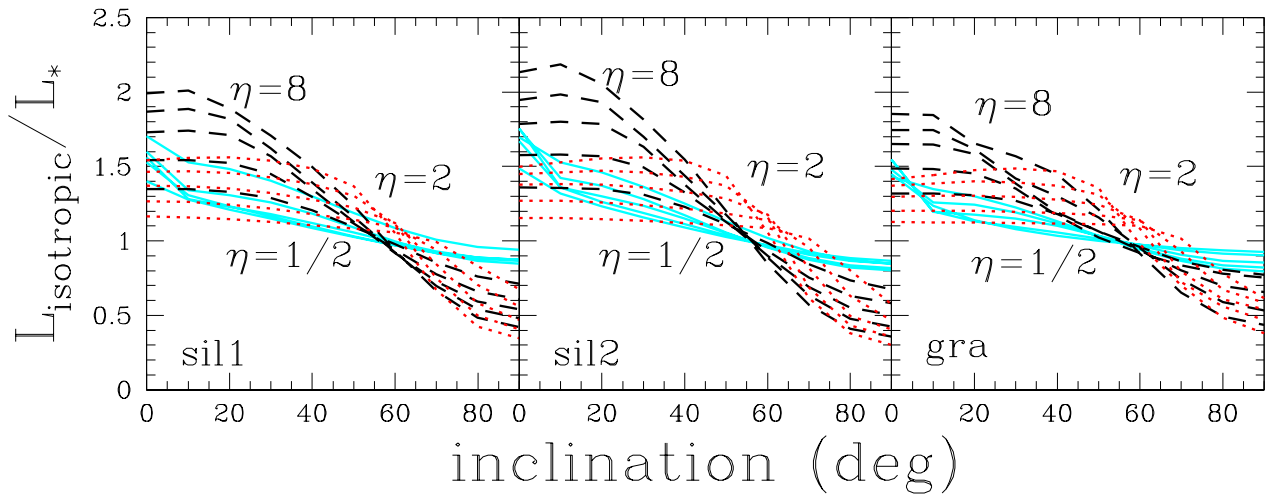}
\caption{
  The isotropic luminosity $L_{isotropic}$ inferred from the SEDs
  in Figs.~\protect\ref{fig:modfh}-\protect\ref{fig:modf8} as a function
  of viewing inclination angle ($0^\circ$ is pole-on, $90^\circ$ is edge-on)
  relative to the true luminosity $L_*$ for the sil1 (left), sil2 (middle)
  and gra (right) dusts.  Results are shown for $\eta=1/2$ (cyan solid),
  $\eta=2$ (black dashed) and $\eta=8$ (red dotted) and optical depths
  of $\tau_e=10$ (least dependence on inclination), $30$, $100$, $300$ 
  and $1000$ (strongest dependence on inclination) for fixed $\tau_p=1$.
  Even for a $\tau_e=1000$ disk viewed exactly edge on, the isotropic
  luminosity is within a factor of $\sim 2$ of the true luminosity. 
  }
\label{fig:lum}
\end{figure*}

\begin{figure*}
\centering
\includegraphics[width=1.00\textwidth]{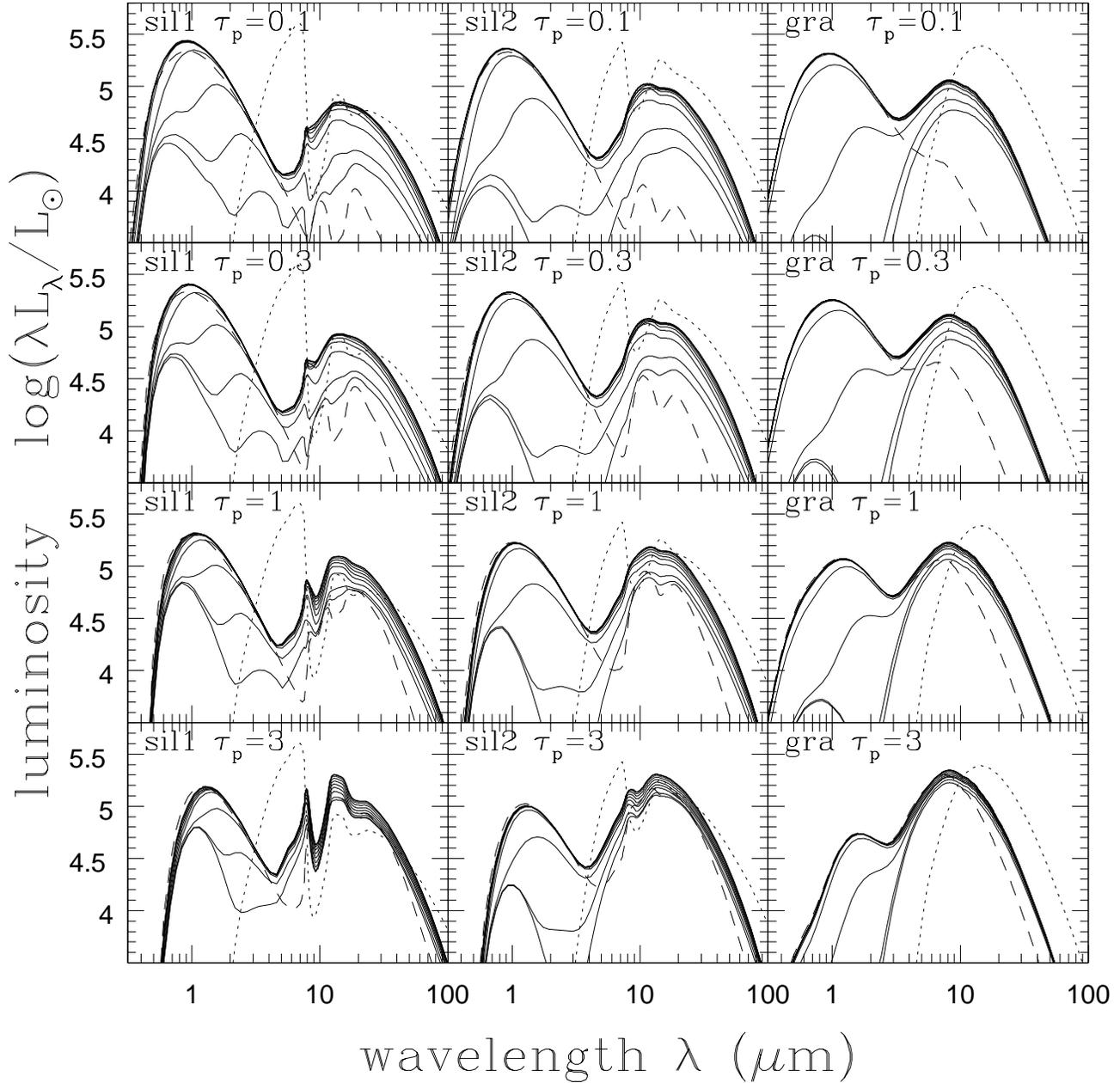}
\caption{
  The effect of varying the polar optical depth over the range $\tau_p=0.1$ (top),
  $0.3$, $1.0$ and $3.0$ (bottom) with $\tau_e=100$, $\eta=8$ and $\Delta t=3$~years
  for the sil1 (left), sil2 (middle) and gra (right) dusts.  The SEDs are again shown for inclinations
  of $0^\circ$ (pole-on) to $90^\circ$ (edge-on) in increments of $10^\circ$ and
  the dotted and dashed lines are for spherical models with optical depths of
  $\tau_e$ and $\tau_p$, respectively. 
  }
\label{fig:pole}
\end{figure*}

\begin{figure*}
\centering
\includegraphics[width=1.00\textwidth]{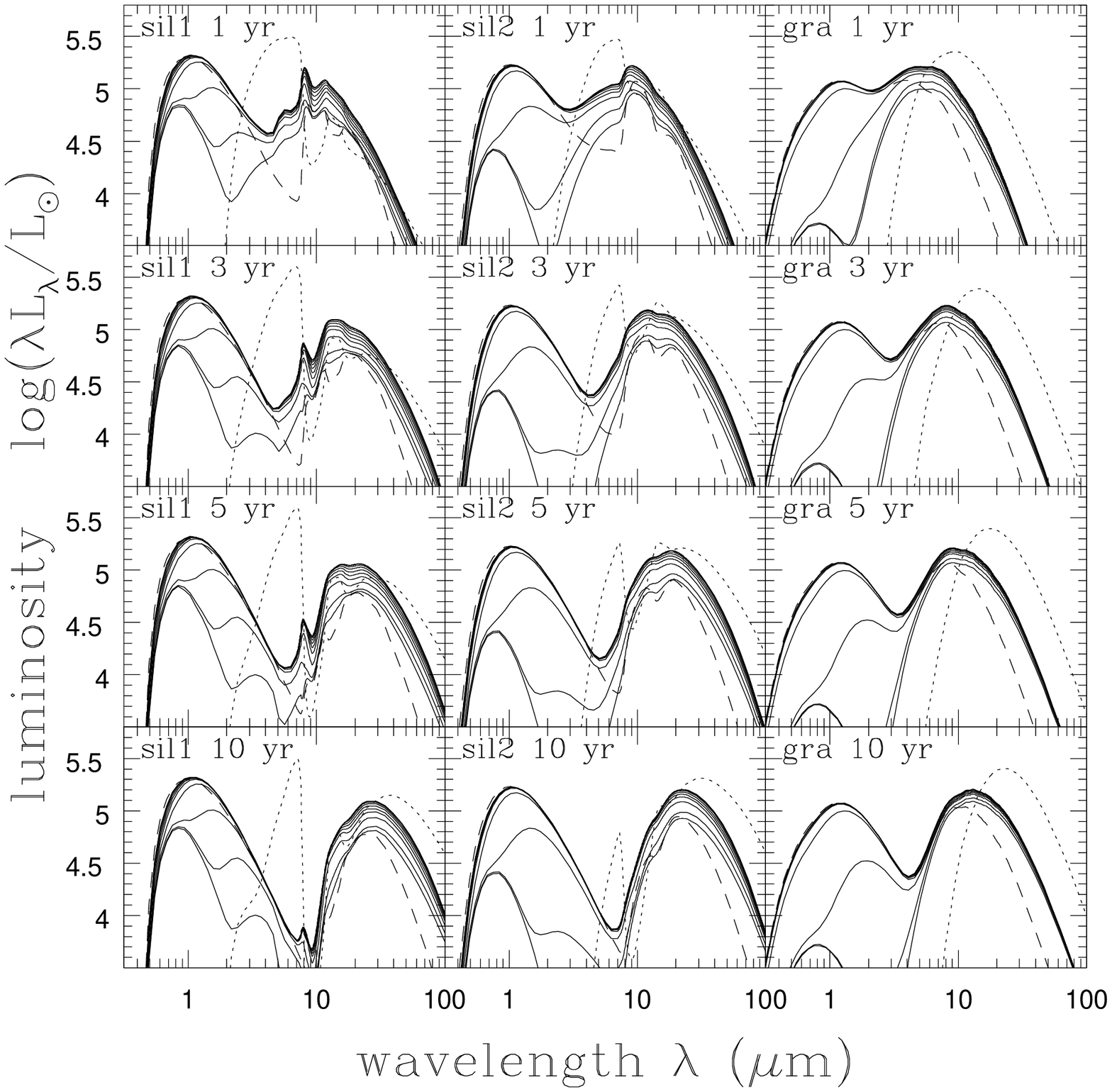}
\caption{
  Inclination dependent SEDs as in Figs.~\protect\ref{fig:modfh}-\protect\ref{fig:modf8}
  for $\eta=8$ and fixed $\tau_e=100$ and $\tau_p=1$ as a function of time, with 
  $\Delta t=1$~year (top), $3$, $5$ and $10$~years (bottom) for the sil1 (left),
  sil2 (middle) and gra (right) dusts.  The SEDs are again shown for inclinations
  of $0^\circ$ (pole-on) to $90^\circ$ (edge-on) in increments of $10^\circ$ and
  the dotted and dashed lines are for spherical models with optical depths of 
  $\tau_e$ and $\tau_p$, respectively.  This is not a real time sequence since
  we are holding the optical depths fixed rather than dropping them as $\Delta t^{-2}$. 
  }
\label{fig:time}
\end{figure*}

\section{Results}

Figs.~\ref{fig:modfh} through \ref{fig:modf8} survey the results as functions
of $\eta$ and $\tau_e$ for fixed $\tau_p=1$ and $\Delta t =3$~yr.  The results
are shown for $\tau_e=10$, $30$, $100$ and $300$ and all three dust types 
with one figure for each of $\eta=1/2$, $1$, $2$, $4$ and $8$, corresponding to an increasingly
disky dust distribution (see Fig.~\ref{fig:density}).  The emergent spectral
energy distribution (SED) is shown for viewing angles from pole-on ($0^\circ$)
to edge on ($90^\circ$) in increments of $10^\circ$, along with the SEDs for 
spherical models with the two limiting optical depths of $\tau_e$ and $\tau_p$

If we focus first on the graphitic models with $\eta=1$ in Fig.~\ref{fig:modf1}
we see that the optical and near-IR emission smoothly shifts between the two
limiting spherical models with viewing angle.  As the equatorial optical depth
increases, the star steadily becomes optically ``invisible'' to higher and higher
latitudes.    The mid-IR emission lies between
the two limiting cases but fairly close to the $\tau_e$ spherical model - for 
example, the spherical models absorb 63\% ($\tau=1)$ and 100\% ($\tau=10$)
of the optical emission, 
respectively, while the disky model with $\tau_e=10$ absorbs 99\% and so the 
total mid-IR emission is very close to the higher optical depth spherical model, although
the dust emission is slightly hotter.  As we raise $\tau_e$, the optical depth
of the disk starts to become important at the wavelengths of the dust emission.
The longer wavelengths continue to simply escape but the shorter wavelengths 
begin to preferentially escape along directions away from the disk.  
Nonetheless, the mid-IR
emission peak is bluer than the high optical depth spherical model from all
viewing directions. 

Two issues make the behavior of the silicate models quite different.  
The first is that the silicate dust has a much higher scattering opacity in 
the optical. Thus, if we look at the optical SEDs for the silicate
models, we always see significantly more optical emission than in the high
optical depth spherical model even when viewed from the equator.  Optical photons heading
towards the polar regions are scattered to observers closer to the equator 
through the reduced optical depth regions higher above the equatorial plane.
This means, however, that the polar optical emission is less than in the low optical
depth spherical model.  The photons scattered to the equatorial observers are not
replaced by initially equatorially directed photons being scattered towards the 
poles because they are absorbed in the disk.

The second difference comes from the strong mid-IR absorption features discussed earlier 
(Fig.~\ref{fig:opdepth}).  In the low optical depth spherical models,
the effects of the absorption features are modest.  However, in the
high optical depth models, the structure of the SED is the inverse of the
opacity, with emission peaking on the blue side of the absorption 
feature, a strong dip in emission at the opacity peak near $10\mu$m, 
a weaker emission peak at the dip in the opacity near $20\mu$m and
then a smoother decline at longer wavelengths.  As the equatorial
optical depth increases, the emission on the blue side of the 
mid-IR opacity peak strengthens while the emission on the red
side remains similar to the spherical model. 

We can understand this as a version of the greenhouse effect created
by the structure of the silicate opacity in Fig.~\ref{fig:opdepth}.
The opacity has a tremendous drop on the blue side of the peak near
$10\mu$m, but
only a slow tail-off on the red side.  So if the dust has a temperature
corresponding to wavelengths around the opacity peak, and the optical
depth is high, then photons bluer than $10\mu$m  escape 
relatively easily while redder photons are reabsorbed. To reach a
radiative equilibrium, the dust heats up until it can radiate 
enough energy in the opacity trough to balance the heating.  As
the optical depth increases, the absorption of redder photons increases
faster than that of the bluer photons, so the emission in the
opacity trough increases relative to the longer wavelength emission.
The emission at $\sim 8\mu$m becomes steadily stronger as the equatorial
opacity increases even when viewed from the equator.

For the less disky $\eta=1/2$ models in Figs.~\ref{fig:modfh}, the
SEDs simply shift to more closely resemble the high optical
depth spherical models.  The polar optical emission is reduced, as is the scattered 
light contribution to more equatorial observers.  The silicate emission
peak also becomes stronger and more closely resembles the spherical
models.

As we now consider the diskier models in Figs.~\ref{fig:modf2} through
\ref{fig:modf8} there are two primary changes.  First, it becomes 
easier to scatter optical photons to an equatorial observer.  Once
a photon emitted upwards is scattered, a thinner disk has less optical depth 
for reabsorbing it when it is scattered towards an equatorial
observer.  For the same equatorial optical depth, there is also simply
more escaping optical light because the concentration of the dust towards
the equator means there is less absorption at intermediate latitudes.
For example, averaging $\exp(-\tau)$ over inclination
for $\tau_p=1$ and $\tau_e=100$, the fraction of V band
photons escaping is $7.5 \times 10^{-5}$, $0.004$, $0.033$, $0.11$
and $0.20$ for $\eta=1/2$, $1$, $2$, $4$ and $8$, respectively.

Second,
the optical depth difference between moving out through the disk versus
perpendicular to the disk becomes increasingly large.  For our angular
structure in the limit that $\tau_p\rightarrow 0$ and we extend the
dust distribution to $R_{out} \rightarrow \infty$, the ratio of the optical
depth from the equator perpendicular to the disk or radially through the
disk is $0.83$ for $\eta=1/2$ and drops to 
$0.57$, $0.35$, $0.21$ and $0.11$
for $\eta=1$, $2$, $4$ and $8$, respectively. This will lead to more of 
the mid-IR emission escaping towards the poles compared to the $\eta=1/2$
model.  For the graphitic dust, this makes the SED shift bluewards 
away from the high optical depth spherical model and closer to the 
lower optical depth spherical model.  For the silicate dusts, the
strength of the features created by the silicate opacity bump weakens
and the SEDs resemble those of a more disky (higher $\eta$) and
lower equatorial optical depth model.  

Fig.~\ref{fig:lum} shows the inferred isotropic luminosities for the
$\eta=1/2$ (Fig.~\ref{fig:modfh}), $\eta=2$ (Fig.~\ref{fig:modf2}) 
and $\eta=8$ (Fig.~\ref{fig:modf8}) as a function of viewing angle
and equatorial optical depths of $\tau_e=10$, $30$, $100$, $300$
and $1000$ with $\tau_p=1$ and $\Delta t=3$~years.  By isotropic
luminosity we mean integrating over the observed SED and assuming
the source would have the same SED if viewed from any direction,
and we compare it to the true luminosity $L_*$.  As can roughly be
inferred from the SEDs, the isotropic luminosity is generally 
fairly similar to the true luminosity.  Viewed pole-on, the isotropic luminosity
is higher because the observer receives both significant optical emission
and infrared emission from the disk.  Viewed edge-on, the isotropic
luminosity is lower because there is little or no optical emission and
the infrared emission is modestly reduced.  But even for the very
disky ($\eta=8$) models with very high optical depths ($\tau_e=1000$), 
the isotropic luminosities are only off from the true luminosities by
factors of $\sim 2$.  The variation with inclination then declines for
less disky configurations (lower $\eta$) or lower equatorial optical
depths.  This gibes with the illuminated slab models considered in
\cite{Adams2017}, which found that even with no emission able to 
escape to the observer through lower optical depth, higher latitude
regions above the disk, the isotropic luminosity was only 
reduced by a factor of $\sim 4$.

Fig.~\ref{fig:pole} shows the consequences of changing the polar
extinction over the range $\tau_p=0.1$, $0.3$, $1.0$ and $3.0$
for $\tau_e=100$, $\eta=8$ and $\Delta t=3$~years. The optical
emission of the graphitic model is simply increasingly suppressed.
Because of the greater importance of scattering for the silicate
the equatorial optical emission initially increases with $\tau_p$
even as the polar emission decreases.  The increasing polar 
optical depth also drives a strengthening of the mid-IR silicate
emission feature, similar to the effects from reducing $\eta$.
Only when the polar optical depth becomes sufficiently high
does the optical equatorial emission begin to drop again.

Finally, in Fig.~\ref{fig:time} we show how the SEDs of the
$\eta=8$ models with $\tau_e=100$ and $\tau_p=1$ depend on the
dust radius, phrased as observation times of $\Delta t=1$,
$3$, $5$ and $10$~years assuming an expansion velocity of
$v_e=10^3$~km/s.  This is not a true time sequence because
we are holding the optical depths fixed.  If we started the
sequence with $\tau_e=100$ at $\Delta t=1$~year, the
optical depths would be $\tau_e=11$, $4$ and $1$
at the later times.  To actually have $\tau_e=100$
at $\Delta t =10$~years, the optical depth at
$\Delta t=1$~year would have to be $\tau_e=10^4$!

Not surprisingly, the optical and near-IR emission depends
little on the radius to the dust.  The geometry of absorbing
and scattering these photons is essentially self-similar.
The mid-IR emission changes because the flux heating the
dust is dropping.  The SED peak shifts to longer wavelengths,
roughly like the black body prediction of $\lambda \propto \Delta t^{1/2}$,
and the silicate emission peak weakens.  The development of
the deep minimum in the SED at $5$-$10\mu$m would make it
relatively easy for these models to evade detection by warm
SST.  This is particularly true for the silicate models, where
the high optical depth spherical models all produce an emission
peak at these wavelengths.  The mid-IR emission peak still lies
at wavelengths easily probed by JWST.   

\section{Discussion}

Disky dust distributions have been invoked in a number of 
contexts to try to better hide stars posited to have survived
an eruptive transient from detection (e.g., \citealt{Kashi2017},
\citealt{Andrews2021}, \citealt{Bear2022}).  They have the
advantage that the require less mass and energy to maintain
a significant optical depth towards an equatorial observer
at late times than a spherical distribution of the same 
optical depth.  It was also generally assumed that a large
fraction of the dust emission which would have been sent towards
an equatorial observer in a spherical geometry would instead 
escape towards the poles, making the source significantly dimmer 
in the mid-IR.

Unless the dust distribution is carefully arranged, using a
disk geometry will tend to enhance the optical emission seen
by an equatorial observer. Photons scatter off dust above
the disk and then can propagate to an equatorial observer.
This can be minimized by having no modest optical depth paths
from the source to the observer -- either no polar dust, or 
enough polar dust that optical emission perpendicular to the
disk is also heavily absorbed.   This is especially true for
silicate dusts with their higher optical scattering opacities.

For silicate dusts, a disk geometry can significantly reduce
the emission at $5$-$10\mu$m  compared to spherical models.
This is a consequence of the strong $10\mu$m silicate dust
absorption feature, which drives a significant emission 
peak on the blue side of the absorption in the spherical
models.  The feature still exists for the disky models, but 
it can be considerably weaker and becomes weaker for diskier 
geometries and colder dust.
The changes in the mid-IR emission for graphitic dust are 
much smaller because the opacity simply declines monotonically
to longer wavelengths.  The mid-IR emission beyond $10\mu$m 
is shifted to bluer wavelengths than a spherical model with 
the same optical depth as the disk.   

None of the models strongly suppress the equatorial flux 
compared to a spherical model.  The isotropic luminosities
of the disk models are only suppressed in the equatorial
direction by a factor of $\sim 2$ even when the optical 
depth through the disk is $\tau_e = 10^3$.  Similarly, the
polar isotropic luminosities are only enhanced by 
similarly modest factors.  \cite{Adams2017} had already
noted that even an infinite slab of dust with similar
optical depths was only reducing the isotropic luminosity
by a factor of 4 despite having no low optical depth 
paths between the observer and the source.

It was always possible to hide stellar luminosity sources
at optical, near-IR and warm-SST mid-IR, which could only
characterize the emission bluewards of $5\mu$m, given 
sufficient dust at a large enough radius.  The low 
resolution of SST also made it difficult to separate
stellar and diffuse emission. The velocities 
required were 
plausible.  The problem was that expansion inexorably 
reduces the optical depth and so the veil either had to
clear on $\sim 10$~year time scales or the mass and energy
budget required to maintain it became un-physical.  
The disk geometry can reduce the mass and energy
requirements to maintain the absorption at later times,
but they must be quite geometrically thin to produce
a large drop in either.

It is essentially impossible to hide survivors of relatively
nearby transients like SN~1997bs, SN~2008S or NGC~6946-BH1
from JWST.  With excellent point source sensitivity out to
$25\mu$m, the mid-IR emission peak predicted in either the
spherical or disky models is observable for decades.  Putting
the dust at a large enough distance to put the peak emission well
beyond $25\mu$m requires impossible masses, kinetic 
energies and elapsed times.

\section*{Acknowledgments}

CSK is supported by NSF grant AST-1908570.

\section*{Data Availability Statement}

All results can be replicated using the publicly available
software package {\tt RADMC-3D} (\citealt{Dullemond2012}).

\end{document}